\documentstyle[prl,aps,epsfig]{revtex}         
\begin{document}
\draft               
\twocolumn[\hsize\textwidth\columnwidth\hsize\csname @twocolumnfalse\endcsname

\title{Patterns and localized structures in bistable semiconductor resonators}
\author{V.B.Taranenko, I.Ganne,$^*$ R.J.Kuszelewicz,$^*$ C.O.Weiss}
\address{Physikalisch-Technische Bundesanstalt 38116 Braunschweig/Germany\\
$^*$Centre National d'Etudes de Telecommunication, Bagneux/France}
\maketitle
\begin{abstract}
We report experiments on spatial switching dynamics and steady
state structures of passive nonlinear semiconductor resonators of
large Fresnel number. Extended patterns and switching front
dynamics are observed and investigated. Evidence of localization
of structures is given.
\end{abstract}
 \vskip1pc ]
It has recently become apparent that pattern formation in optics
is related in many ways with other fields of physics \cite{tag:1}.
One simple optical pattern forming system is a nonlinear
resonator, the subject of recent investigations see, e.g.
\cite{tag:2} and op.cit. The analogy of resonator optics with
fluids \cite{tag:3}, particularly, suggests a variety of phenomena
not considered for optics before \cite{tag:4}. For active
resonators (lasers, lasers with nonlinear absorber, 4-wave mixing
oscillators) predicted phenomena, such as vortices and spatial
solitons, have already been demonstrated experimentally
\cite{tag:2}. On the other hand, experimental results for passive
resonators, recently studied extensively theoretically with a view
to possible application \cite{tag:5,tag:6}, are limited to the
early demonstration of structures in resonators containing liquid
crystals or alkali vapours as the nonlinear medium \cite{tag:7}.
We report here the first experimental investigations of structure
formation in passive nonlinear semiconductor resonators. These
systems, apart from possible usefulness in applications, show
phenomena in optics analogous to those found in other fields of
nonlinear physics (e.g. optical Turing instability \cite{tag:8} or
competition between pattern formation and switching \cite{tag:9}).
Our observations include regular (hexagonal) pattern formation
and, in presence of optical bistability (OB), space- and
time-resolved switching waves. We also present observations which
indicate mutual locking of OB switching waves to form localized
structures (spatial solitons) \cite{tag:15}. Further evidence of
localization effects is shown by independently switch bright spots
of a hexagonal structure. \\

The resonator used for the experiment consists of GaAs/GaAlAs
multiple quantum well (MQW) material (18 periods of
GaAs/Ga$_{0.5}$Al$_{0.5}$As with 10 nm/10 nm thickness) between
Bragg mirrors of about 99.5 $\%$ reflectivity on a GaAs substrate.
Properties of these microresonators have been described in
\cite{tag:10}. The optical resonator length is approximately 3
$\mu$m with a corresponding free spectral range of 50 THz. The
resonator thickness varies over the usable sample area (10 x 20
mm). So, by choosing a particular position on the sample, it is
possible to vary the cavity resonance wavelength such that its
downshift $\Delta \lambda$ below the exciton line lies in the
range 0 $<$ $\Delta \lambda$ $<$ 25 nm. \\ The absorption of the
MQW material depends on $\Delta \lambda$. A typical finesse of the
resonator far below the exciton energy is 200, which corresponds
to 125 GHz half width at half maximum (HWHM) of the cavity
resonance. \\ The radiation source used is a continuous-wave
Ar$^+$-pumped Ti:Al$_2$0$_3$-laser. To realize a reasonable
Fresnel number, the radiation has a spot width on the sample of
about 60 $\mu$m. The nonlinearity used is predominantly dispersive
and defocusing. The characteristic (bistable/monostable) and shape
of the resonator response depends on both $\Delta \lambda$ and the
detuning $\delta \lambda$ of the laser field from the resonator
resonance. \\ As the substrate for the resonator is opaque, all
observations are done in reflection. The OB describing the
reflection returned from the input surface is "N-shaped"
\cite{tag:10}, and complementary to the intracavity field
characteristic, which has the familiar "S-shape". To avoid
confusion, we will therefore refer to "switched" and "unswitched",
rather then "upper" and "lower", states. \\ Within the illuminated
area, smaller areas ($\approx$ 8 $\mu$m) can be irradiated by
short pulses ($\leq$ 0.1 $\mu$s) to initiate local switching. The
light of these "injection" pulses is polarized perpendicularly to
and "incoherent" with the "background-" or "holding-" beam.
Address pulses thus produce an injection of photocarriers, locally
changing the optical properties of the resonator. \\ All
observations are made within times lasting a few microseconds,
repeated at 1 kHz, to eliminate thermal effects. Acousto-optic
modulators (AOM) serve for fast intensity modulation of the
injected optical fields, with a time resolution of 50 ns. They can
be programmed to produce complex pulse-shapes, and to synchronize
the drive and address pulses. \\ A variety of detection equipment
is employed. A CCD camera records two-dimensional images, but with
slow time response. Images are thus most useful for stable
steady-state structures. Observations of intrapulse dynamics are
done with a fast detector (2 ns), which monitors the incident
power, and measures reflected light intensity with a spatial
resolution of 4 $\mu$m. The good repeatability of the
spatio-temporal dynamics allows to map the intensity dynamics on a
diameter of the illuminated area, by successively imaging the
points of the diameter onto the detector while recording the
intensity in each point as a function of time.
\\

At large $\Delta \lambda$ and small negative $\delta \lambda$
($\mid$$\delta \lambda$$\mid$ $\leq$ 1 HWHM) hexagonal structures
form (Fig. 1). Hexagonal patterns appear in many different fields
of physics \cite{tag:11}, and have also been predicted, but not
previously observed, for nonlinear resonators
\cite{tag:12,tag:13}. The period of the hexagonal lattice is about
20 $\mu$m, consistent with typical predictions for semiconductor
models \cite{tag:13}. Our interpretation of Fig. 1 is that it is
the result of a supercritical modulational instability of the
unswitched branch. There is no observable threshold intensity for
this pattern, however. We attribute this to a blurring of the
threshold due to spatial and temporal variations and fluctuations
of the input beam. Note that Fig. 1 is the reflected signal, so
the imaged negative hexagons (lattice of dark spots) corresponds
in terms of intracavity intensity to positive hexagons (lattice of
bright spots). \\
\begin{figure}[htbf]
\epsfxsize=60mm \centerline{\epsfbox{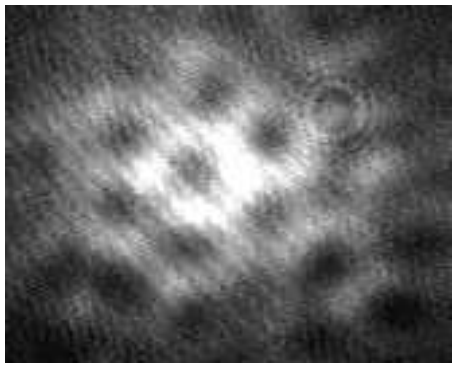}} \vspace{0.5cm}
\caption{Hexagonal pattern observed for the following parameters:
$\Delta \lambda$ = 25 nm, $\delta \lambda$ = -1 HWHM, power 30 mW,
size of illuminated area 60 $\mu$m.}
\end{figure}
The lattice period is measured to scale linearly with
$1/\sqrt{\delta \lambda}$ as expected for tilted waves
\cite{tag:12,tag:14}, which we expected to be the basic mechanism
of this hexagon formation. We observe a change from negative to
positive hexagons with decreasing $\delta \lambda$. The contrast
of the pattern decreases with increasing $\delta \lambda$. Above
$\delta \lambda$ = 1.5 HWHM OB switching occurs before a pattern
with notable contrast develops, when increasing the intensity. \\
Since the illumination is with a Gaussian beam, the central part
of the field switches first. The switched domain is separated from
the surrounding unswitched area by a stationary switching front
[15]. Such a switching front moves if the incident light intensity
is different from the "Maxwellian" \cite{tag:15} intensity, the
intensity for which the potential maxima for lower and upper
branch are equally deep. The front moves into the unswitched area
if the local incident intensity is larger than the Maxwellian
intensity \cite{tag:15} and vice versa. It is stationary only on
the contour on which the incident light intensity equals the
Maxwellian intensity. Thus switching fronts can be moved by
changing the incident light intensity, as Fig. 2 shows. \\ Fig. 2a
gives the incident and reflected intensity at the center of the
Gaussian beam as a function of time; 2b and 2c show the intensity
along a diameter of incident and reflected beam respectively as a
function of time in the form of equiintensity contours. The
recording Fig. 2d gives the reflectivity of the resonator along
the same diameter of the Gaussian beam. \\
\begin{figure}[htbf]
\epsfxsize=70mm \centerline{\epsfbox{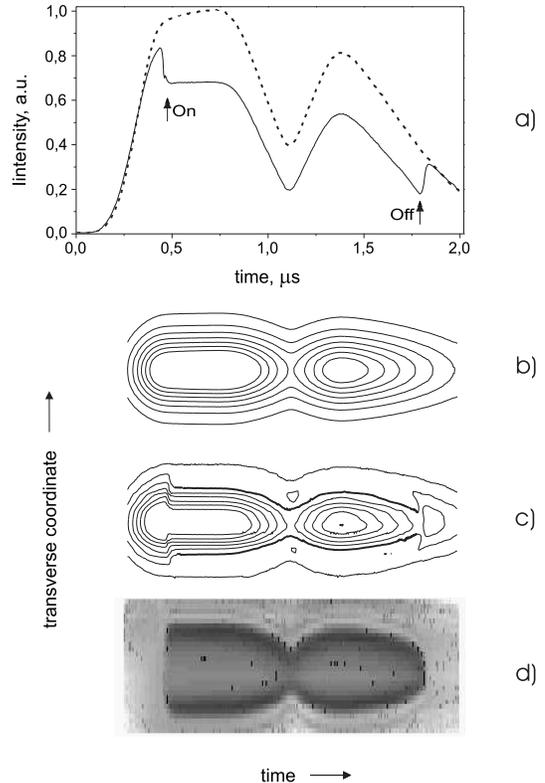}} \vspace{0.5cm}
\caption{Motion of switching front: a)  incident light intensity
(dashed) and light intensity reflected from sample, center of
Gaussian beam (solid), b)  intensity contours of incident light,
c) intensity contours of reflected light (switching front matches
the third lowest intensity contour of incident light), d)
reflectivity of sample (darkest zones are the switching front).\\
Parameters are: $\Delta \lambda$ = 17 nm, $\delta \lambda$ = -1.6
HWHM, power 70 mW, spot size 60 $\mu$m.\\ Same time scale for a,
b, c, d.}
\end{figure}
The light intensity is programmed here by the AOM as described
above in order to study switching-wave dynamics. An initial peak
(2a) switches most of the beam cross section to the low
reflectivity state. During the rapid initial outward motion of the
switching front (as well as the inward motion at switch-off) the
position of the switching front is probably not adiabatically
controlled by the light field. As predicted \cite{tag:15}, for
adiabatic motion the switching front follows an intensity contour
of the varying incident light; as evident by comparing 2b and 2c.
\\ As opposed to the extended patterns (Fig. 1) localized
structures (spatial cavity solitons) have been predicted
\cite{tag:5,tag:6}. Such structures can form due to interaction
between switching fronts, in which context they have been called
"diffractive autosolitons" \cite{tag:15}.\\
\begin{figure}[htbf]
\epsfxsize=70mm \centerline{\epsfbox{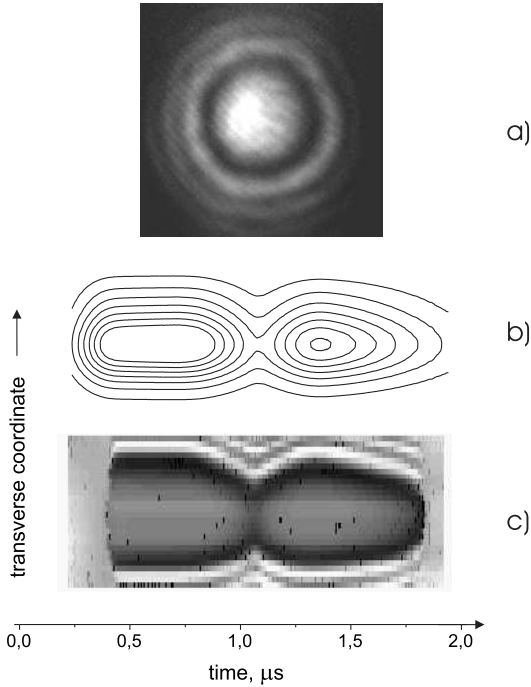}} \vspace{0.5cm}
\caption{Switching front with oscillating tails: a)  switching
front (darkest ring) surrounded by equidistant rings in unswitched
area, b) intensity contours of incident light, c) reflectivity of
sample.\\ Parameters are:  $\Delta \lambda$ = 25 nm, $\delta
\lambda$ = -1.6 HWHM, power 80 mW, spot size 60 $\mu$m.}
\end{figure}
If switching fronts have "oscillating tails", the gradients
associated with these oscillations can mutually trap two switching
fronts. In 2D such "oscillating tails", which can occur due to a
nearby modulational instability \cite{tag:16}, may stabilize a
circular switching front. This is equivalent to a localized
structure or spatial soliton, which is free to move as a whole
unless constrained by boundary effects. This mechanism of
formation of such spatial solitons was studied theoretically
\cite{tag:17} and demonstrated already experimentally
\cite{tag:18} on a system with phase bistability. Oscillating
tails are readily observable in the present system by choosing
appropriate $\delta \lambda$ and $\Delta \lambda$; as seen in Fig.
3. Similar front-locking may thus occur in systems with intensity
bistability such as the present one. The stabilization of front
distances can thus be used as one criterion for the existence of
spatial solitons. Another criterion is evidently the moveability
of localized structures, which, furthermore, implies bistability
of the structure. \\ Fig. 4 shows a corresponding observation. The
intensity on a diameter of the illuminated field is recorded as a
function of time. The background illumination is set in the middle
of the bistability range.\\
\begin{figure}[htbf]
\epsfxsize=60mm \centerline{\epsfbox{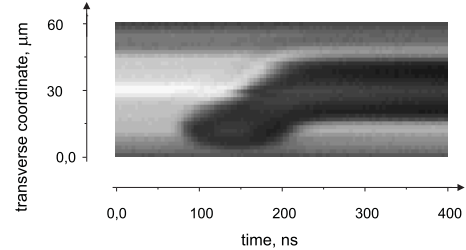}} \vspace{0.5cm}
\caption{Motion of switching front for off-center local
switching.\\ Parameters of the holding beam as in Fig. 3.}
\end{figure}
At some distance from the beam center a narrow (8 $\mu$m)
injection pulse can locally switch a small area to the upswitched
branch. Fig. 4 shows how this switched spot moves to the center of
the illuminated area (due to the intensity gradient of the
background field). It would appear that the two points of the
circular switching front, which can be followed in Fig. 4, move in
parallel, suggesting a moving stable structure. \\
\begin{figure}[htbf]
\epsfxsize=40mm \centerline{\epsfbox{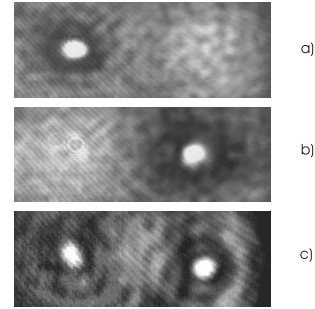}} \vspace{0.5cm}
\caption{Location of bright switched spot on two intensity maxima
of the background field.\\ Observation 400 $\mu$m below sample
surface (see text).}
\end{figure}
After the structure has reached the center of the field, where the
intensity is maximal, it remains there stationarily. When there
are two or more local intensity maxima in the background
illumination, the final position of the up-switched structure can
be at any of the maxima. Fig. 5 shows a background illumination,
which has an intensity saddle at the center of the picture and
local maxima at the center of the left and right half of the
picture. Depending on whether the injection is (anywhere) in the
right or the left half of the picture, the final position of the
up-switched structure will be the left or right intensity maximum
respectively (Fig. 5a,b). Equally, at the two maxima up-switched
structures can exist simultaneously (Fig. 5c).\\
\begin{figure}[htbf]
\epsfxsize=70mm \centerline{\epsfbox{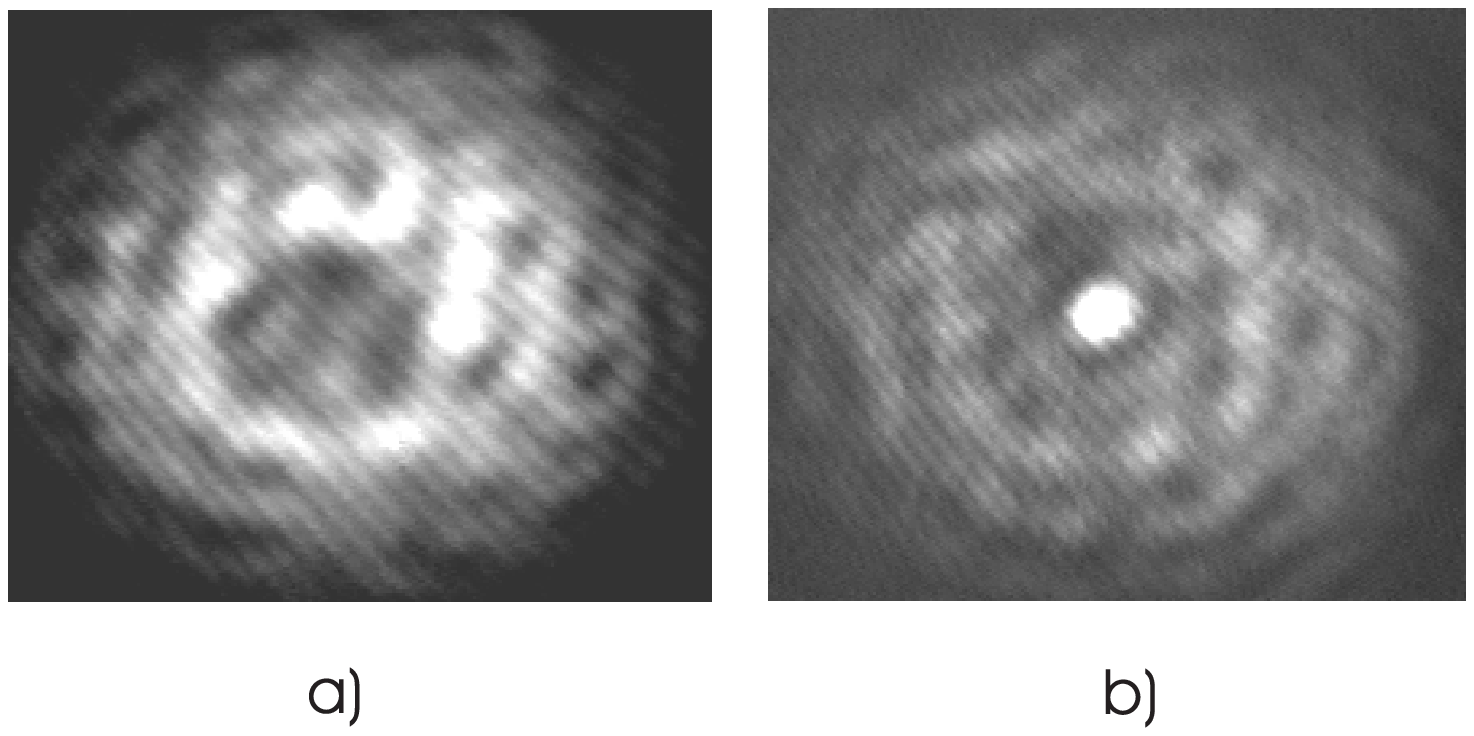}} \vspace{0.5cm}
\caption{Switched spot embedded in structured background.
Observation on (a)) and 400 $\mu$m below (b)) sample surface. \\
Parameters are:  $\Delta \lambda$ = 25 nm, $\delta \lambda$ = -1
HWHM power 40 mW, spot size 60 $\mu$m.}
\end{figure}
At small $\delta \lambda$, when the background intensity is
increased beyond the appearance of a high contrast hexagonal
pattern a noticeably small spot ($\approx$ 10 $\mu$m) appears as
the switched-up structure, Fig. 6. For better measurement signal
to noise ratio we have recorded in Figs 5 to 8 in a plane 400
$\mu$m below the sample surface. In this plane the structures
appear bright, regular, with high contrast compared to the
structures in the sample plane. Thus allowing to reduce the
recording averaging times.\\
\begin{figure}[htbf]
\epsfxsize=70mm \centerline{\epsfbox{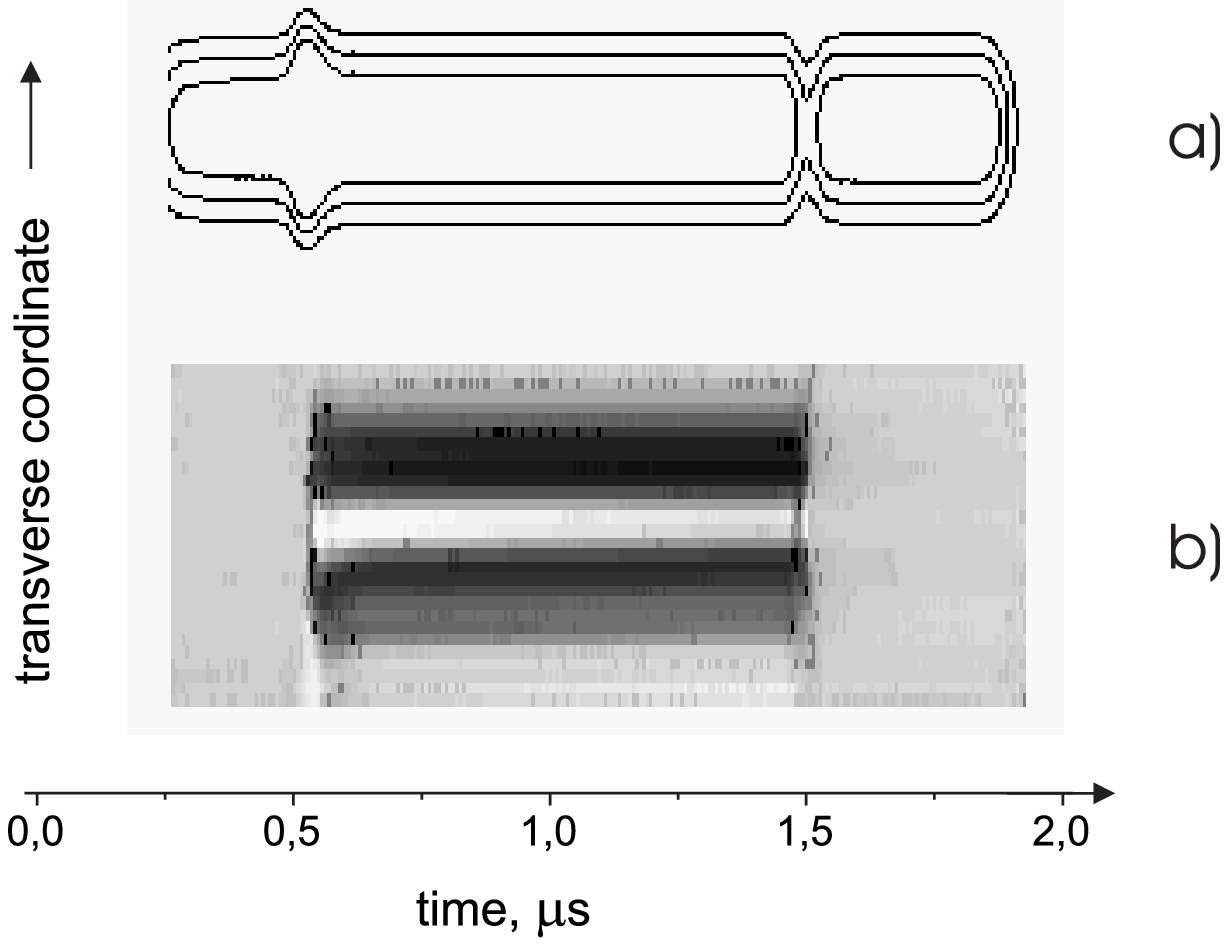}} \vspace{0.5cm}
\caption{Bistability of localized structure. a)  intensity
contours of incident light, b)  reflectivity of sample.\\ Fast
short increase of incident light intensity switches localized
structure on. Fast short reduction of incident light intensity
switches localized structure off.\\ Parameters as in Fig. 6.}
\end{figure}
Because of the small size and high brightness of the structure in
Fig. 6b further tests were done on its soliton properties. The
first test (bistability) is shown in Fig. 7. Fig. 7a shows
intensity contours of the incident light in the same way as Fig.
2b. Fig. 7b shows the light intensity reflected from the sample.
The initial background intensity is chosen in the middle of the
bistability range. A short (pulsed) increase of intensity (at ca.
0.5 $\mu$s) is then seen to switch the sample up. The small bright
structure remains up-switched when the intensity is returned to
its initial value below the switching limit. The small structure
is switched back off by momentary decrease of the illumination to
below the lower bistability limit (at ca. 1.5 $\mu$s). This is
clear proof of bistability as expected for a spatial soliton. \\
\begin{figure}[htbf]
\epsfxsize=70mm \centerline{\epsfbox{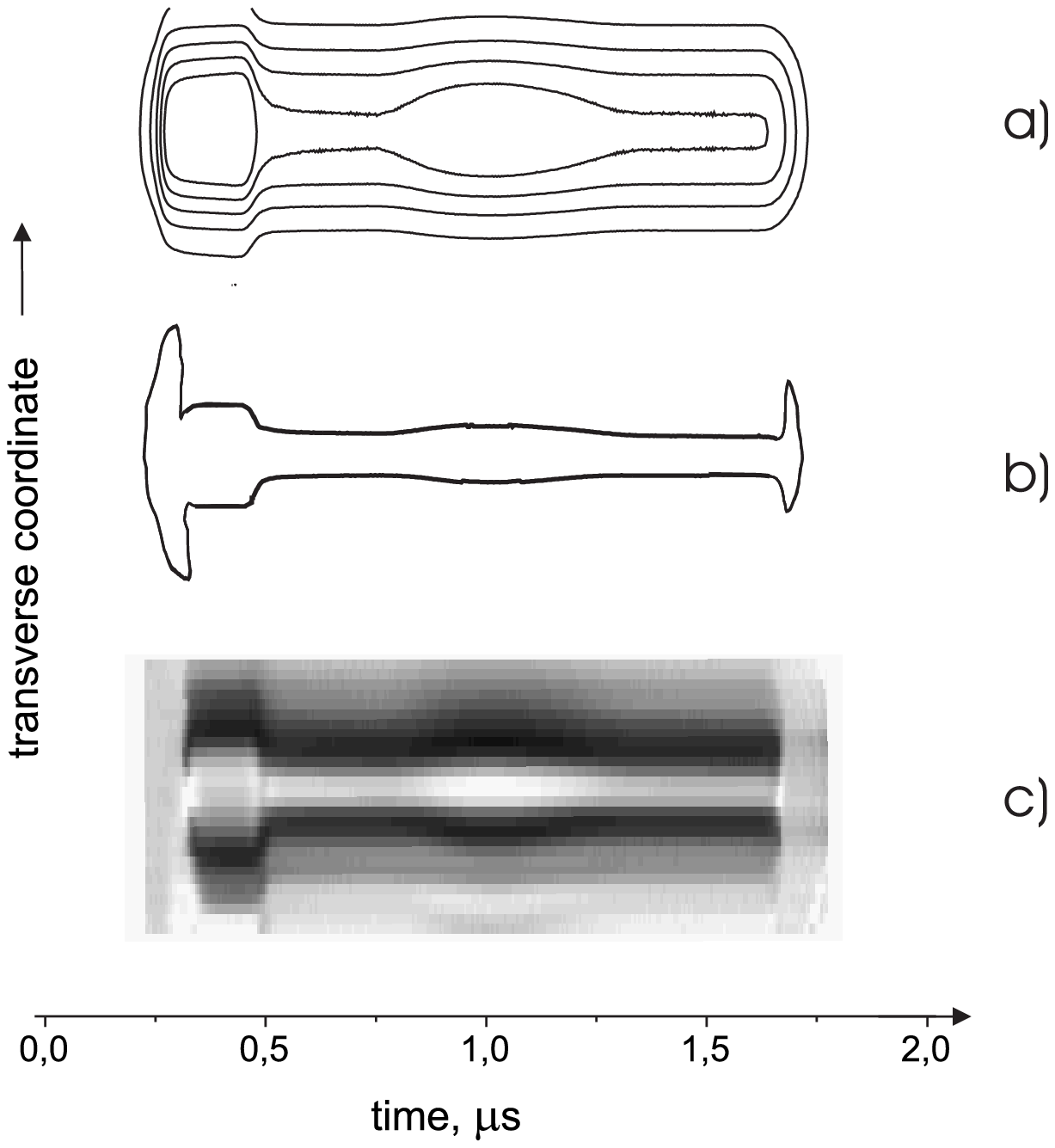}} \vspace{0.5cm}
\caption{Test of localized structure robustness. a)  intensity
contours of incident light, b)  intensity contour of reflected
light, c)  reflectivity of sample.\\ Note that there is no
intensity contour of the incident light which matches the contour
of the bright zone.\\ Parameters as in Fig. 6.}
\end{figure}
If this structure is just a small up-switched domain, then the
switching fronts surrounding it should follow an equiintensity
contour of the background field. If, on the other hand, there is
"locking" of the switching front, it should not follow precisely
the intensity changes. The result of the test is shown in Fig. 8:
8a and 8b give intensity contours for the incident and reflected
light respectively, 8c gives the reflectivity. At the beginning
the small structure is created by an initial pulse above the
switching threshold, followed by a reduction of input intensity to
a value in the middle of the bistability range. The test for
robustness is then done by variation of the intensity (within the
bistability range). Fig. 8b,c shows that the spatial variation of
the reflected intensity, and thus of the switching front, is much
less marked than that of the incident light intensity, in contrast
to what we observe in Fig. 2. This indicates a "robustness" of the
structure indicative of self-localization. \\ Fig. 9 gives
evidence that in hexagonal patterns the individual bright spots
can have properties of localized structures or spatial solitons.
9a shows a hexagonal structure seemingly similar to Fig. 1. The
intensity of the background field is in the middle of the
bistability range. Injection with a narrow beam in a short pulse
aimed at the bright spot marked "a" in 6a, switches it, as seen in
9b, to be a dark "defect". When the injection beam is aimed at the
adjacent spot (marked "b"), this adjacent spot is switched off,
all other spots remaining unchanged. To demonstrate that the
switched spot is stable, we show in Fig. 9d the output from that
region as a function of time during the 2 $\mu$s duration of the
main input pulse. In the upper trace, the address pulse is too
weak to induce switching, and the output recovers within 100ns to
its original steady value. In the lower trace, switching does
occur, and the output remains almost constant at a level less than
20 $\%$ below its original value, until the holding light is
returned to zero. Thus unlike a coherent extended hexagon pattern,
the structure of Fig. 9a behaves like a collection of localized
structures, who can independently be switched. \\
\begin{figure}[htbf]
\epsfxsize=80mm \centerline{\epsfbox{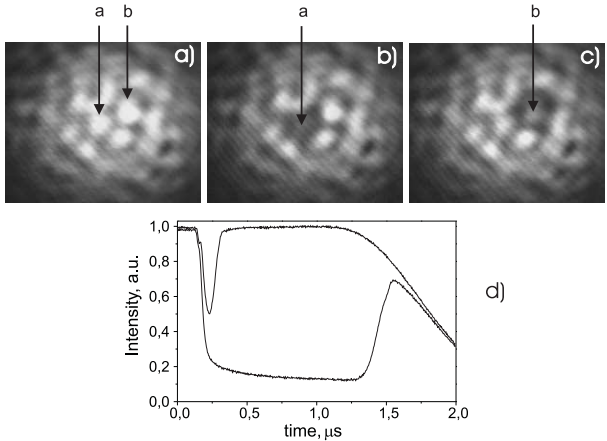}} \vspace{0.5cm}
\caption{Switching of individual bright spots of a bright spot
claster (see text).\\ Parameters are: $\Delta \lambda$ = 25 nm,
$\delta \lambda$ = -1 HWHM, holding beam width 60 $\mu$m,
injecting beam width 8 $\mu$m.}
\end{figure}

Concluding, we have shown for the first time the formation of
hexagon patterns, as predicted for non-linear passive resonators.
Evidence was given of soliton properties of small spatial
structures. \\

Acknowledgements \\ This work was supported by ESPRIT LTR project
PIANOS. Discussions with the project partners are acknowledged. We
thank W.J.Firth for valuable suggestions. We also thank
K.Staliunas for developing clarifying concepts such as the linear
filtering of spatial noise by a 2D resonator. \\

\end{document}